\documentclass[letterpaper]{article} 
\usepackage[submission]{aaai23}  
\usepackage{times}  
\usepackage{helvet}  
\usepackage{courier}  
\usepackage[hyphens]{url}  
\usepackage{graphicx} 
\usepackage{natbib}  
\usepackage{caption} 
\usepackage{algorithm}
\usepackage{algorithmic}

\usepackage{newfloat}
\usepackage{listings}
\usepackage{balance}

\usepackage[switch]{lineno}

\usepackage{times}

\usepackage{soul}
\usepackage{url}
\usepackage[utf8]{inputenc}
\usepackage{graphicx}
\usepackage{amsmath}
\usepackage{booktabs}
\usepackage{subcaption}
\usepackage{xspace}
\usepackage{xcolor}
\usepackage{footnote}
\usepackage{amssymb}
\usepackage{pifont}
\newcommand{\cmark}{\ding{51}}%
\newcommand{\xmark}{\ding{55}}%

\newcommand{\csr}{committee selection rule\xspace}

\newcommand{\DiReCFp}[2]{DRCF\xspace}
\newcommand{\DiReCWD}{DRCWD\xspace}
\newcommand{\DiReCWDp}[2]{DRCWD\xspace}
\newcommand{\DiReCWDpse}[2]{DRCWD\xspace}
\newcommand{\DiReCWDpsu}[2]{DRCWD\xspace}
\newcommand{\DiRe}{DiRe\xspace}
\newcommand{\DIRE}{DiRe\xspace}
\newcommand{\group}[2]{A_{(#1, #2)}}
\newcommand{\groupsub}[2]{A_{(#1, #2_#1)}}
\newcommand{\population}[2]{P_{(#1, #2)}}
\newcommand{\populationsub}[2]{P_{(#1, #2_#1)}}

\newcommand{\etal}{\emph{et~al.}\xspace}
\usepackage{amsthm}
\newtheorem{definition}{Definition}
\newtheorem{theorem}{Theorem}
\newtheorem{example}{Example}
\newtheorem{corollary}{Corollary}

\usepackage{multirow}
\usepackage{xfrac}
\usepackage{subcaption}
\usepackage{balance}
\usepackage[switch]{lineno} %

\usepackage{imakeidx}
\usepackage{mathtools}

\DeclareMathOperator{\pos}{pos}

\DeclareMathOperator{\WU}{WU}
\DeclareMathOperator{\U}{U}
\DeclareMathOperator{\Borda}{Borda}

\DeclareCaptionStyle{ruled}{labelfont=normalfont,labelsep=colon,strut=off} 
\floatstyle{ruled}
\newfloat{listing}{tb}{lst}{}
\floatname{listing}{Listing}
\title{My Publication Title --- Single Author}
\title{Fairly Allocating Utility in Fair Multiwinner Elections}
\title{Fairly Allocating Utility in Constrained Multiwinner Elections}
\author {
    Kunal Relia\thanks{This work was supported, in part, by Julia Stoyanovich's NSF grants No. 1934464 and 1916505.}\\
    {\normalfont New York University, USA}\\
    {\normalfont krelia@nyu.edu}
}

\begin{document}
\maketitle

\begin{abstract}
Fairness in multiwinner elections is studied in varying contexts. For instance, diversity of candidates, representation of voters, or both are separately termed as being fair. A common denominator to ensure fairness across all such contexts is the use of constraints. However, across these contexts, the candidates selected to satisfy the given constraints may systematically lead to unfair outcomes for historically disadvantaged voter populations as the \emph{cost of fairness} may be borne unequally. Hence, we develop a model to select candidates that satisfy the constraints fairly across voter populations. To do so, the model maps the constrained multiwinner election problem to a problem of fairly allocating indivisible goods. We propose three variants of the model, namely, \emph{global}, \emph{localized}, and \emph{inter-sectional}. Next, we analyze the model's computational complexity and present an empirical analysis of the utility traded-off across various settings of our model. 
We observe the potential impact of Simpson's paradox on results using synthetic datasets and a dataset of voting at the United Nations. 
Finally, we discuss the implications of our work on studies that use constraints to guarantee fairness.

\end{abstract}

\section{Introduction}
\label{sec:intro}

Fairness is receiving particular attention from the computer science research community. 
Specifically, there is a growing trend among the Algorithmic Game Theory and the  Computational Social Choice communities toward the use of ``fairness'' \cite{bredereck2018multiwinner,celis2017multiwinner,cheng2019group,flanigan2021fair,hershkowitz2021district,relia2021dire,shrestha2019fairness}. 
Moreover, the term is used in varying contexts. For example, Celis \etal \shortcite{celis2017multiwinner} call diversity of candidates in committee elections fairness, Cheng \etal \shortcite{cheng2019group} call representation of voters in committee elections fairness, and Relia \shortcite{relia2021dire} call diversity of candidates \emph{and} representation of voters in committee elections fairness. 
A common denominator across all such papers that guarantee fairness is the use of constraints.
For example, Celis \etal \shortcite{celis2017multiwinner} use diversity constraints to be fair to candidate groups 
and Cheng \etal \shortcite{cheng2019group} use representation constraints to be fair to voter populations. Relia \shortcite{relia2021dire} unified these frameworks to select a diverse and representative (\DiRe) committee.


\begin{figure}[t!]
\centering
\begin{subfigure}{.5\textwidth}
  \centering
  \includegraphics[width=.5\textwidth]{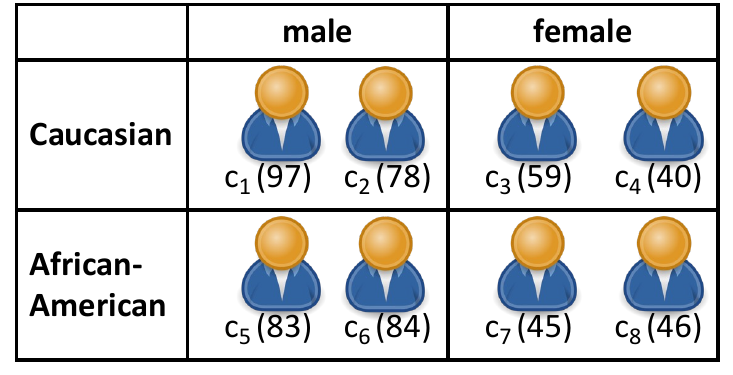}
  \caption{\textbf{candidates}}
  \label{fig:whatexample/candidates}
\end{subfigure}%
\newline
\begin{subfigure}{.5\textwidth}
  \centering
  \includegraphics[width=.8\textwidth]{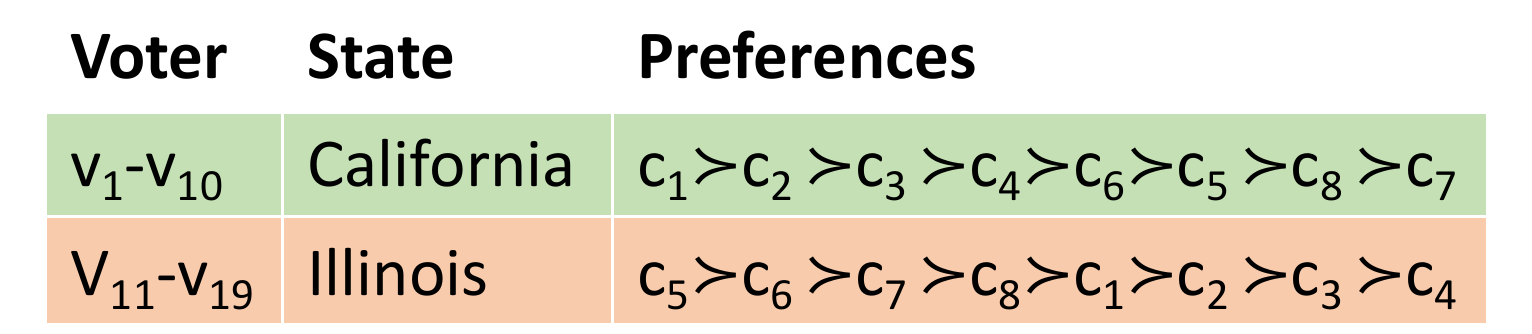}
  \caption{\textbf{voters}}
  \label{fig:whatexample/voters}
\end{subfigure}%
\caption{(a) Candidates with ``race'' and ``gender'' attributes (and their Borda scores). (b) Voters with ``state'' attribute. The winning committee (size $k$=4) for California and Illinois, states in the United States, is \{$c_1,c_2,c_3,c_4\}$ and \{$c_5,c_6,c_7,c_8\}$, respectively.}
\label{fig:whatexample}
\end{figure}

\begin{example}
\label{eg-dire}
Consider an election $E$ consisting of $m =8 $ candidates and $n=19$ voters giving ordered preference over $m$ candidates to select a committee of size $k=4$. Each candidate has two attributes, race and gender (Figure~\ref{fig:whatexample/candidates}) (e.g., candidate $c_1$ is a Caucasian male). Voters have one attribute, state (Figure~\ref{fig:whatexample/voters}) (e.g., voters $v_1$ to $v_{10}$ belong to the state of California). The $k$-Borda\footnote{The Borda rule associates score $m - i$ with the $i^{th}$ position, and $k$-Borda selects the candidates with the $k$ highest Borda scores.} winning committee computed for each voter population is  $\{c_1,c_2,c_3,c_4\}$ for California and $\{c_5,c_6,c_7,c_8\}$ for Illinois.
The candidates that form an optimal committee of size $k=4$ consists of $W=\{c_1$, $c_2$, $c_6$, $c_5\}$ with a committee score, $\mathtt{f}(W)=342$. 

\DIRE Committee enforces \emph{diversity constraint} that requires the committee to have at least two candidates of each gender and each race, and a \emph{representation constraint} that requires at least two candidates from the winning committee of each state. Observe that the optimal committee\footnote{For simplicity, the example uses $k$-Borda instead of Chamberlin-Courant rule (discussed in Section~\ref{sec:prelim}). Even when the latter guarantees proportional representation, our motivation of fairly allocating utility holds.}, which is also representative, consists of $W=\{c_1, c_2, c_5, c_6\}$ ($\mathtt{f}(W)=342$), but this committee is not diverse, since all candidates are male. The highest-scoring \DiRe committee is $W'=\{c_1, c_6, c_3, c_8\}$ ($\mathtt{f}(W')=286$).
\end{example}



A \DiRe committee ensures fairness in terms of the number of candidates getting selected from candidate groups and voter populations. However, an unintended consequence of enforcing constraints is that it may be fair to various voter populations in a systematically unequal way. Hence, it is important to assess \emph{who} pays \emph{what} cost of fairness. This assessment can be done in multiple ways depending on the context so as to ensure that we fairly allocate fairness across all the voter populations. 

\begin{example}
\label{eg-balanced}
The highest-scoring \DiRe committee selected in Example~\ref{eg-dire} was $W'=\{c_1, c_6, c_3, c_8\}$ ($\mathtt{f}(W')=286$). Note that this outcome fails to select Illinois' highest-ranked candidate ($c_5$), but selects California's highest-ranked candidate ($c_1$). Additionally, if we calculate the total utility derived from the committee by each state, the smaller state achieves lower utility as compared to the larger state (14 versus 16). In contrast, if we select $W''=\{c_1, c_5, c_3, c_8\}$ ($\mathtt{f}(W'')=285$), then both the states get their most preferred candidate and have equal utility (15), at a small cost of the total committee utility.
\end{example}

This example illustrates two techniques\footnote{Similar techniques are well-studied in fair allocation of indivisible goods \cite{freemanrecent}.} that can be used to fairly allocate utility in a \DiRe multiwinner election: (i) selecting the most favorite candidate and (ii) equating utilities received from the winning committee by each population. In either case, without our proposed mitigation techniques, the \DiRe committee is unequally fair across various populations, specifically, harming smaller, historically under-represented populations. 
An important observation we make here is that even if the attributes of the candidates and of the voters \emph{coincide}, we need to treat them separately as fairness to candidates may still cause unfairness to voters. For instance, consider that the voter attribute in Figure~\ref{fig:whatexample/voters} was gender instead of the state. Hence, voters $v_1$ to $v_{10}$ are male and voters $v_{11}$ to $v_{19}$ are female. Therefore, based on Example~\ref{eg-balanced},
this change implies that the candidates selected to satisfy the constraints that require a female candidate on the committee systematically led to unequal fairness for female voters, which should not be the case. 


\paragraph{Global, Localized, and Inter-sectional Fair Allocation:} Global fair allocation refers to fair allocation of utility across \emph{all} populations under \emph{all} voter attributes and localized fair allocation refers to fair allocation of utility across \emph{similar} populations under the \emph{same} voter attribute. If there are $\pi$ attributes, then we either do one global fair allocation or $\pi$ localized fair allocation. Localized fair allocation may be needed especially when each voter has more than one attribute. This is because fairly allocating utility between male voters and African-American voters, for example, is not realistic. Fair allocation of utility to male voters should be compared with a voter population under the gender attribute only. On the other hand, inter-sectional fair allocation refers to fair allocation of utility across inter-sectional populations. For example, Caucasian males, Caucasian females, African-American males, and African-American females.

\noindent{\bf Contributions:} 
\begin{itemize}
    \item We develop a model that uses various techniques for fairly allocating candidates across populations in a constrained multiwinner election to mitigate unequal fairness caused among the voter populations.
    \item We propose three variants of the model, namely, global, localized, and inter-sectional.
    \item We study the model theoretically and empirically, and 
    show the impact of Simpson's paradox between global / localized fair allocation and inter-sectional fair allocation using synthetic and real-world datasets.
\end{itemize}


\section{Related Work}
\label{sec:RW}


Our work primarily builds upon the literature on constrained multiwinner elections. Goalbase score functions, which specify an arbitrary set of logic constraints and let the score capture the number of constraints satisfied, could be used to ensure diversity \cite{uckelman2009representing}. Using diversity constraints over multiple attributes in single-winner elections is NP-hard \cite{lang2018multi}. Also, using diversity constraints over multiple attributes in multiwinner elections and participatory budgeting is NP-hard, which has led to approximation algorithms and matching hardness of approximation results by Bredereck \etal \shortcite{bredereck2018multiwinner} and Celis \etal \shortcite{celis2017multiwinner}. Finally, due to the hardness of using diversity constraints over multiple attributes in approval-based multiwinner elections \cite{brams1990constrained}, these have been formalized as integer linear programs (ILP) \cite{potthoff1990use}. In contrast, Skowron \etal \shortcite{skowron2015achieving} showed that ILP-based algorithms fail in the real world when using ranked voting-related unconstrained proportional representation rules. 

Overall, the work by Bredereck \etal \shortcite{bredereck2018multiwinner}, Celis \etal \shortcite{celis2017multiwinner}, Relia \shortcite{relia2021dire}, Suhr \etal \shortcite{suhr2019two}, and Yang \etal \shortcite{yang2019balanced} is closest to ours but we differ as we: (i) propose a model that can fairly allocate fairness in varying contexts, and (ii) consider the consequence of enforcing fairness on one or more actors of the system to one main actor of a single-sided platform. Additionally, our work also differs from Conitzer \etal \shortcite{conitzer2019group} as we use predefined populations and use each population's collective preferences. 

\section{Preliminaries and Notation}
\label{sec:prelim}


\paragraph{Multiwinner Elections.} Let $E = (C, V )$ be an election consisting of a candidate set $C = \{c_1,\dots,c_m\}$ and a voter set $V = \{v_1,\dots,v_n\}$, where each voter $v \in V$ has a preference list $\succ_{v}$ over $m$ candidates, ranking all of the candidates from the most to the least desired. $\pos_{v}(c)$ denotes the position of candidate $c \in C$ in the ranking of voter $v \in V$, where the most preferred candidate has position 1 and the least preferred has position $m$. 

Given an election $E = (C,V)$ and a positive integer $k \in [m]$ (for $k \in \mathbb{N}$, $[k] = \{1, \dots, k\}$), a multiwinner election 
selects a $k$-sized subset of candidates (or a committee) $W$ using a multiwinner voting rule $\mathtt{f}$ (discussed later) such that the score of the committee $\mathtt{f}(W)$ is the highest. 
We assume ties are broken using a pre-decided priority order.

\paragraph{Candidate Groups.} 
The candidates have $\mu$ attributes, $A_1 , . . . , A_\mu$, such that $\mu \in \mathbb{Z}$ and $\mu \geq 0$. Each attribute $A_i$, $\forall$ $i \in [\mu]$, partitions the candidates into $g_i \in [m]$ groups, $\group{i}{1} , . . . , \groupsub{i}{g} \subseteq C$. Formally, $\group{i}{j} \cap  \group{i}{j'} = \emptyset$, $\forall j,j' \in [g_i], j \ne j'$. For example, candidates in Figure~\ref{fig:whatexample/candidates} have race and gender attribute  ($\mu$ = 2) with disjoint groups per attribute, male and female ($g_1$ = 2) and Caucasian and African-American ($g_2$ = 2). Overall, the set $\mathcal{G}$ of \emph{all} such arbitrary and potentially non-disjoint groups is $\group{1}{1} , . . . , \groupsub{\mu}{g} \subseteq C$. 



\paragraph{Voter Populations.}
The voters have $\pi$ attributes, $A'_1 , . . . , A'_\pi$, such that $\pi \in \mathbb{Z}$ and $\pi \geq 0$. The voter attributes may be different from the candidate attributes. Each attribute $A'_i$, $\forall$ $i \in [\pi]$, partitions the voters into $p_i \in [n]$ populations, $\population{i}{1} , . . . , \populationsub{i}{p} \subseteq V$. Formally, $\population{i}{j} \cap  \population{i}{j'} = \emptyset$, $\forall j,j' \in [p_i], j \ne j'$. For example, voters in Figure~\ref{fig:whatexample/voters} have state attribute  ($\pi$ = 1), which has populations California and Illinois ($p_1$ = 2). Overall, the set $\mathcal{P}$ of \emph{all} such predefined and potentially non-disjoint populations will be $\population{1}{1} , . . . , \populationsub{\pi}{p} \subseteq V$. 

Additionally, we are given $W_P$, the winning committee $\forall$ $P \in \mathcal{P}$. We limit the scope of $W_P$ to be a committee instead of a ranking of $k$ candidates because when a \csr such as CC rule is used to determine each population’s winning committee $W_P$, then a complete ranking of each population’s collective preferences is not possible. 


\paragraph{Multiwinner Voting Rules.} There are multiple types of multiwinner voting rules, also called committee selection rules. 
In this paper, we focus on committee selection rules $\mathtt{f}$ that are based on single-winner positional voting rules, and are
monotone and submodular ($\forall A \subseteq B, f(A) \leq f(B)$ and $f(B) \leq f(A) + f(B \setminus  A)$).


\begin{definition}
\label{def-CC} \textbf{Chamberlin–Courant (CC) rule \cite{chamberlin1983representative}:}
The CC rule  associates each voter with a candidate in the committee who is their most preferred candidate in that committee. 
 The score of a committee is the sum of scores given by voters to their associated candidate. 
 Specifically, $\mathbf{\beta}$-CC uses the Borda positional voting rule such that it assigns a score of $m-i$ to the $i^{\text{th}}$ ranked candidate who is their highest-ranked candidate in the committee.

        
\end{definition}
A special case of submodular functions are separable functions: 
score of a committee $W$ is the sum of the scores of individual candidates in the committee. Formally, $\mathtt{f}$ is separable if it is submodular and $\mathtt{f}(W) = \sum_{c \in W}^{}\mathtt{f}(c)$ \cite{bredereck2018multiwinner}. 
Monotone and separable selection rules are natural and are considered good when the goal of an election is to shortlist a set of individually excellent candidates:


\begin{definition}
\label{def-kborda} \textbf{$k$-Borda rule} The $k$-Borda rule
outputs committees of $k$ candidates with the highest Borda scores.
\end{definition}

Note that we focus on fairly allocating candidates \emph{only} across voter populations. This is because monotone, submodular scoring functions like $\beta$-Chamberlin Courant do not give scores to individual candidates. Hence, there is no way to fairly allocate candidates across candidates groups as these rules score a committee and not each candidate. We used $k$-Borda in Examples ~\ref{eg-dire} and \ref{eg-balanced} for simplicity.

\section{Fair Allocation Model}
\label{sec:Model}

In this section, we formally define a model that maps the \DIRE Committee Winner Determination problem to a problem of fairly allocating indivisible goods. The model mitigates unfairness to the voter population caused by \DIRE Committees. We first define constraints. 
\vspace{-0.0175cm}
\paragraph{Diversity Constraints,} denoted by $l^D_G \in [1,$ $\min(k, |G|)]$ 
for each candidate group $G \in \mathcal{G}$, enforces at least $l^D_G$ candidates from the group $G$ to be in the committee $W$. Formally,  $\forall$ $G \in \mathcal{G}$, $|G \cap W|\geq l^D_G$. 

\vspace{-0.0175cm}
\paragraph{Representation Constraints,} denoted by $l^R_P \in [1,k]$ for each voter population $P \in \mathcal{P}$, enforces at least $l^R_P$ candidates from the population $P$'s committee $W_P$ to be in the committee $W$. Formally,  $\forall$ $P \in \mathcal{P}$, $|W_P \cap W|\geq l^R_P$.

\begin{definition}
\label{def-DiReCWD}
\textbf{\DIRE Committee Winner Determination (\DiReCWD):}  We are given an instance of election $E=(C,V)$,  
a committee size $k \in [m]$, a set of candidate groups $\mathcal{G}$ under $\mu$ attributes and their diversity constraints $l^D_G$  $\forall$ $G \in \mathcal{G}$, a set of voter populations $\mathcal{P}$ under $\pi$ attributes and their representation constraints $l^R_P$ and the winning committees $W_P$  $\forall$ $P \in \mathcal{P}$, and a \csr $\mathtt{f}$.
Let $\mathcal{W}$ denote the family of all size-$k$ committees. 
The goal of \DiReCWD is to select committees $W \in \mathcal{W}$ that satisfy  diversity and representation constraints such that $|G\cap W|\geq l^D_G$  $\forall$ $G\in \mathcal{G}$ and $|W_P\cap W|\geq l^R_P$  $\forall$ $P\in \mathcal{P}$ and maximizes, $\forall W \in \mathcal{W}$, $\mathtt{f}(W)$. Committees that satisfy the constraints are \DiRe committees.
\end{definition}

Example~\ref{eg-balanced} showed that a \DiRe committee may create or propagate biases by systematically increasing the selection of lower preferred candidates. This may result in more loss in utility to historically disadvantaged voter populations.

To mitigate this, we model our problem of selecting a committee as allocating goods (candidates) into $k$ slots. Hence, to assess the quality of candidates being selected from voter populations, we borrow ideas from the literature on the problem of fair resource allocation of indivisible goods \cite{brandt2016handbook}. Formally, given a set of $n$ agents, a set of $m$ resources, and a valuation each agent gives to each resource, the problem of fair allocation is to partition the resources among agents such that the allocation is fair. There are three classic fairness desiderata, namely, proportionality, envy-freeness, and equitability \cite{freemanrecent}. Intuitively, proportionality requires that each agent should receive her ``fair share'' of the utility, envy-freeness requires no agent should wish to swap her allocation with another agent, and equitability requires all agents should have the exact same value for their allocations and no agent should be jealous of what another agent receives. As balancing the loss in utility to candidate groups is analogous to balancing the fairness in ranking \cite{yang2019balanced}, we focus on balancing the loss in utility to voter populations. We propose varying notions of envyfreeness to balance the loss in utility to voter populations\footnote{Our model, \emph{Fairly Allocating} Utility in Constrained Multiwinner Elections, is analogous to a Swiss Army knife. The model is applicable to any context of a constrained multiwinner election and the setting of the model can be chosen based on the context.}:

\begin{table*}[t]
\centering
\begin{tabular}{l|l||c|c||c|c||c|c|c||c|c|c}
\toprule
 & Committee & score & DiRe & \multicolumn{2}{c||}{FEC} & \multicolumn{3}{c||}{UEC} & \multicolumn{3}{c}{WEC}\\
& &  & Committee & \multicolumn{2}{c||}{up to} & \multicolumn{3}{c||}{up to} & \multicolumn{3}{c}{up to}\\
 \cline{5-12}
&  &  &  & 0 & 1 & 0 & 1 & 2 & 0 & $\frac{1}{13}$& $\frac{2}{13}$\\
\midrule
1&$\{c_1, c_2, c_6, c_5\}$&342& \xmark &\cmark&\cmark&\cmark&\cmark&\cmark&\cmark&\cmark&\cmark\\
\hline

2&$\{c_1, c_6, c_3, c_8\}$&286& \cmark &\xmark&\cmark&\xmark&\xmark&\cmark&\xmark&\xmark&\cmark\\
\hline

3&$\{c_1, c_5, c_3, c_8\}$&285& \cmark &\cmark&\cmark&\cmark&\cmark&\cmark&\xmark&\cmark&\cmark\\
\hline

4&$\{c_1, c_5, c_3, c_7\}$&284& \cmark &\cmark&\cmark&\xmark&\xmark&\cmark&\cmark&\cmark&\cmark\\

\bottomrule
\end{tabular}
\caption{Properties satisfied by various example committees selected using the election setup given in Figure~\ref{fig:whatexample/candidates} and Figure~\ref{fig:whatexample/voters}. Each row corresponds to a committee being (row 1) optimal, (row 2) optimal DiRe, (row 3) optimal FEC and optimal UEC given it is DiRe, and (row 4) optimal WEC given it is DiRe. `DiRe' denotes a committee satisfying diversity and representation constraints. `FEC' denotes Favorite-Envyfree-Committee. `UEC' denotes Utility-Envyfree-Committee. `WEC' denotes Weighted-Envyfree-Committee.}

\label{tab:FECUECWEC-examplesummary}
\end{table*}

 \subsection{Favorite-Envyfree-Committee (FEC)} 
 Each population deserves their top-ranked candidate to be selected in the winning committee. However, selecting a \DiRe committee may result into an imbalance in the position of the most-favorite candidate selected from each population's ranking. A natural relaxation of FEC is finding a committee with a bounded level of envy. Specifically, in the relaxation of FEC up to $x$ where $x \in [k]$, rather than selecting the most preferred candidate, we allow for one of the top-$(x+1)$ candidates to be selected. Note that when $x=0$, FEC and FEC up to 0 are equivalent.

Note that the relaxation, FEC up to $x$ is useful when a FEC does not exist. If a FEC exists, then FEC up to $x$ exists for all non-negative $x$.

 \subsection{Utility-Envyfree-Committee (UEC)} 




Each population deserves to minimize the difference between the utilities each one gets from the selected winning committee, where the utility is the sum of Borda scores that the population gives to the candidates in the winning committee. However, a \DiRe committee may result into an unequal utility amongst all the populations. Formally, for each $P \in \mathcal{P}$, the utility that the population gets from a winning \DiRe committee $W$ is:
    \begin{equation} \U_P = \sum_{c \in  W} \mathtt{f}_{\Borda}^{W_P}(c) \end{equation}
where $\mathtt{f}_{Borda}^{W_P}(c)$ is the Borda score that candidate $c$ gets based on its rank in population $P$'s winning committee $W_P$. Overall, a UEC is a $k$-sized committee such that it aims to

\begin{equation} \forall P, P' \in \mathcal{P} : P \neq P', |\U_P - \U_{P'}| = 0 \end{equation}


A natural relaxation of UEC is UEC up to $\eta$ where $\eta \in \mathbb{Z}: \eta \in [0,$ $\frac{(m-1)\cdot(m)}{2}]$. This is to say that each population deserves to minimize the difference between utilities from the selected winning committee but up to $\eta$. Hence, a UEC  up to $\eta$ is a $k$-sized committee such that it aims to

\begin{equation} \forall P, P' \in \mathcal{P} : P \neq P', |\U_P - \U_{P'}| \leq \eta \end{equation}

Note that in line with FEC, the relaxation, UEC up to $\eta$ is useful when a UEC does not exist. If a committee $W$ is UEC, then it implies that $W$ is UEC up to $\eta$ for all $\eta$ in $[0,$ $\frac{(m-1)\cdot(m)}{2}]$. The relation does not hold the other way, which is to say that if a committee $W$ is UEC up to $\eta$, then it may not be UEC up to $\eta-1$.







\subsection{Weighted-Envyfree-Committee (WEC)} Each population deserves to minimize the difference between the weighted utilities they get from the selected winning committee. The weighted utility is the sum of Borda scores that the population gives to the candidates in the winning committee who are among their top-$k$ candidates over the maximum Borda score that they can give to candidates based on their representation constraint. Formally, for each $P \in \mathcal{P}$, the weighted utility that the population having representation constraint $l_P^R$ gets from a winning \DiRe committee $W$ is:
    \begin{equation} \WU_P = \frac{\sum_{c \in W_P \cap W} \mathtt{f}_{\Borda}^{W_P}(c)}{\sum_{i=1}^{l_P^R}m-i} \end{equation}

Overall, a WEC is a $k$-sized committee such that it aims to

\begin{equation} \forall P, P' \in \mathcal{P} : P \neq P', |\WU_P - \WU_{P'}| = 0 \end{equation}

\begin{example}
\label{eg-WECcalc}
The highest-scoring \DiRe committee selected in Example~\ref{eg-dire} was $W'=\{c_1, c_6, c_3, c_8\}$ ($\mathtt{f}(W')=286$). The $\WU_{IL}$ will be calculated as follows: Given, $l_{IL}^R$=2, $W_P \cap W = \{c_6, c_8\}$.

\[\WU_{IL} = \frac{\mathtt{f}_{\Borda}^{W_{IL}}(c_6) + \mathtt{f}_{\Borda}^{W_{IL}}(c_8)}{7+6} = \frac{6+4}{7+6}=\frac{10}{13}\]

Similarly, $\WU_{CA}$ will be calculated as follows: Given, $l_{CA}^R$=2, $W_P \cap W = \{c_1, c_3\}$.

\[\WU_{CA} = \frac{\mathtt{f}_{\Borda}^{W_{CA}}(c_1) + \mathtt{f}_{\Borda}^{W_{CA}}(c_3)}{7+6} = \frac{7+5}{7+6}=\frac{12}{13}\]
\end{example}

A natural relaxation of WEC is WEC up to $\zeta$ where $\zeta \in \mathbb{Q}$ such that $\zeta \in [0,$ $1]$. This is to say that each population deserves to minimize the difference between weighted utilities from the selected winning committee but up to $\zeta$. Hence, a WEC  up to $\zeta$ is a $k$-sized committee such that it aims to

\begin{equation} \forall P, P' \in \mathcal{P} : P \neq P', |\WU_P - \WU_{P'}| \leq \zeta \end{equation}


\section{Complexity Results}
\label{sec:complexity}

In this section, we analyze the computational complexity of various settings of our global fair allocation model. 

Note that global fair allocation is a generalization of localized fair allocation. Hence, a polynomial-time algorithm we give for the former holds for the latter (but with trivial modification). On the other hand, the NP-hardness of the localized fair allocation implies the NP-hardness of the global fair allocation but not vice versa. Hence, we design each NP-hardness reduction for global fair allocation such that it holds for localized and inter-sectional fair allocation as well.

\subsection{FEC} 

We first present a polynomial-time algorithm for Favorite-Envyfree-Committee (FEC), FEC up to $k-1$, and FEC up to $k-2$.

\begin{theorem}\label{thm:FECuptoxP}
Given a winning \DIRE Committee $W$ and an integer $x$ in $\{0,$ $k-2,$ $k-1\}$, there is a polynomial time algorithm that determines whether a FEC up to $x$, W', exists such that for all candidate groups $G$ in $\mathcal{G}$, the diversity constraint $l^D_G \leq |G \cap W'|$ and for all voter populations $P$ in $\mathcal{P}$, the representation constraint $l^R_P \leq |W_P \cap W'|$.
\end{theorem}

\begin{algorithm}[t!]
\caption{FEC Algorithm}
\label{alg:algorithmFEC}
\begin{flushleft}
\textbf{Input}:\\
$W_P$, $\forall P \in \mathcal{P}$ - winning committee of each population\\
$l_P^R$, $\forall P \in \mathcal{P}$ - diversity constraints\\
$l_G^D$, $\forall G \in \mathcal{G}$ - representation constraints\\
$W$ - Winning \DIRE Committee\\
\textbf{Output}: True if FEC exists, False otherwise\\
\end{flushleft}
\begin{algorithmic}[1] 
\IF{$x=0$}
\STATE $S$ = $\bigcup {c\in C}: \forall P \in \mathcal{P},$ $ \max_{c \in W_P} \mathtt{f}_{\Borda}^{W_P}(c)$

\STATE \textbf{if} $|S|>k$ \textbf{then} \textbf{return} {False}
\STATE \textbf{if} $S\subseteq W$ and  $|S|>0$ \textbf{then} \textbf{return} {True}
\FOR{\textbf{each} $G \in \mathcal{G}$}
\STATE \textbf{if} $|G \cap S|<l_G^D$ \textbf{then} \textbf{return} {False}
\ENDFOR
\FOR{\textbf{each} $P \in \mathcal{P}$}
\STATE \textbf{if} $|W_P \cap S|<l_P^R$ \textbf{then} \textbf{return} {False}
\ENDFOR
\STATE \textbf{return} {True}
\ENDIF
\IF{$x=k-1$}
\STATE \textbf{return} {True}
\ENDIF
\IF{$x=k-2$}
\STATE $S$ = $\phi$

\FOR{\textbf{each} $P \in \mathcal{P}$}
\STATE $S$ = $W \setminus$ $\{\forall c \in W_P : c$ is the lowest scoring candidate $\mathtt{f}_{\Borda}^{W_P}(c)$ among $W_P\}$
\FOR{\textbf{each} $P \in \mathcal{P}$}
\STATE \textbf{if} $|W_P \cap S|<l_P^R$ \textbf{then} \textbf{return} {False}
\ENDFOR

\FOR{\textbf{each} $G \in \mathcal{G}$}
\STATE \textbf{if} $|G \cap S|<l_G^D$ \textbf{then} \textbf{return} {False}
\ENDFOR
\ENDFOR
\STATE \textbf{return} {True}
\ENDIF
\end{algorithmic}
\end{algorithm}

\begin{proof}
Algorithm~\ref{alg:algorithmFEC} runs in time polynomial of the size of the input. For correctness, consider the following cases:

\begin{itemize}
    \item \textbf{When $\mathbf{x=0}$}, FEC up to 0 can only exist if $\forall P \in \mathcal{P}$, the population top-scoring candidate $\mathtt{f}_{\Borda}^{W_P}(c)$ is in the committee. Hence, we select each and every top-ranked candidate into set $S$. Then, if set $S$ satisfies all the given constraints, then we know FEC up to 0 exists.
    \item \textbf{When $\mathbf{x=k-1}$}, then the existence of a \DiRe committee implies the existence of FEC as FEC up to $k-1$ is equivalent to satisfying the requirement that each population has at least one candidate in the committee, which is in line with the definition of the \DiRe committees.
    \item \textbf{When $\mathbf{x=k-2}$}, we iterate over each population $P \in \mathcal{P}$ such that we remove the population's winning committee's $W_P$ least favorite candidate. If the candidates that remain in $W$ satisfy all the constraints after the removal of this candidate, then it implies that FEC up to $k-2$ exists. This is because if a \DiRe committee exists even after removing a population's least favorite candidate, then it implies that FEC up to $k-2$ also exists. 
\end{itemize}

\end{proof}

We now present hardness\footnote{The hardness results throughout the paper are under the assumption P $\neq$ NP.} results and differ the proofs to the appendix. 

\begin{theorem}\label{thm:FECuptoxNP-hard}
Given a \DIRE Committee $W$ and an integer $x$ in $[1,$ $k-3]$, it is NP-hard to determine whether a FEC up to $x$, W', exists such that for all candidate groups $G$ in $\mathcal{G}$, the diversity constraint $l^D_G \leq |G \cap W'|$ and for all voter populations $P$ in $\mathcal{P}$, the representation constraint $l^R_P \leq |W_P \cap W'|$, even when  $l^D_G=0$ and $l^R_P=0$.
\end{theorem}

\subsection{UEC} A Utility-Envyfree-Committee (UEC) is a $k$-sized committee such that it aims to

\[ \forall P, P' \in \mathcal{P} : P \neq P', |\U_P - \U_{P'}| = 0 \]

\begin{theorem}\label{thm:UECNP-hard}
Given a \DIRE Committee $W$, it is NP-hard to determine whether a UEC, $W'$, exists such that for all candidate groups $G$ in $\mathcal{G}$, the diversity constraint $l^D_G \leq |G \cap W'|$ and for all voter populations $P$ in $\mathcal{P}$, the representation constraint $l^R_P \leq |W_P \cap W'|$, even when $l^D_G=0$ and $l^R_P=0$.
\end{theorem}

The hardness of UEC up to $\eta$ follows from the hardness of UEC.
\begin{corollary}\label{thm:UECuptoetaNP-hard}
Given a \DIRE Committee, it is NP-hard to determine whether W', a UEC up to $\eta$, for all $\eta$ in $\mathbb{Z}$ such that $\eta$ in $[0,$ $\frac{(m-1)\cdot(m)}{2})$, exists such that for all candidate groups $G$ in $\mathcal{G}$, the diversity constraint $l^D_G \leq |G \cap W'|$ and for all voter populations $P$ in $\mathcal{P}$, the representation constraint $l^R_P \leq |W_P \cap W'|$.
\end{corollary}

Note that when $\eta=0$, UEC up to $\eta$ and UEC are equivalent. When $\eta=\frac{(m-1)\cdot(m)}{2}$, UEC up to $\eta$ and \DiRe committee are equivalent.
    
\subsection{WEC} A Weighted-Envyfree-Committee (WEC) is a $k$-sized committee such that it aims to

\[ \forall P, P' \in \mathcal{P} : P \neq P', |\WU_P - \WU_{P'}| = 0 \]

\begin{theorem}\label{thm:WECNP-hard}
Given a \DIRE Committee, it is NP-hard to determine whether a WEC, $W'$, exists such that for all candidate groups $G$ in $\mathcal{G}$, the diversity constraint $l^D_G \leq |G \cap W'|$ and for all voter populations $P$ in $\mathcal{P}$, the representation constraint $l^R_P \leq |W_P \cap W'|$, even when $l^D_G=0$ and $l^R_P=0$.
\end{theorem}

The hardness of WEC up to $\zeta$ follows from the hardness of WEC.

\begin{corollary}\label{thm:WECuptozetaNP-hard}
Given a \DIRE Committee, it is NP-hard to determine whether W', a WEC up to $\zeta$, for all $\zeta$ in $\mathbb{Q}$ such that $\zeta$ in $[0,$ $1)$, exists such that for all candidate groups $G$ in $\mathcal{G}$, the diversity constraint $l^D_G \leq |G \cap W'|$ and for all voter populations $P$ in $\mathcal{P}$, the representation constraint $l^R_P \leq |W_P \cap W'|$.
\end{corollary}

Note that when $\zeta=0$, WEC up to $\zeta$ and WEC are equivalent. When $\zeta=1$, WEC up to $\zeta$ and \DiRe committee are equivalent.

\begin{figure*}[t!]
\centering
\begin{subfigure}{.23\textwidth}
  \centering
  \includegraphics[width=\linewidth, trim=0cm 0cm 0cm 0cm]{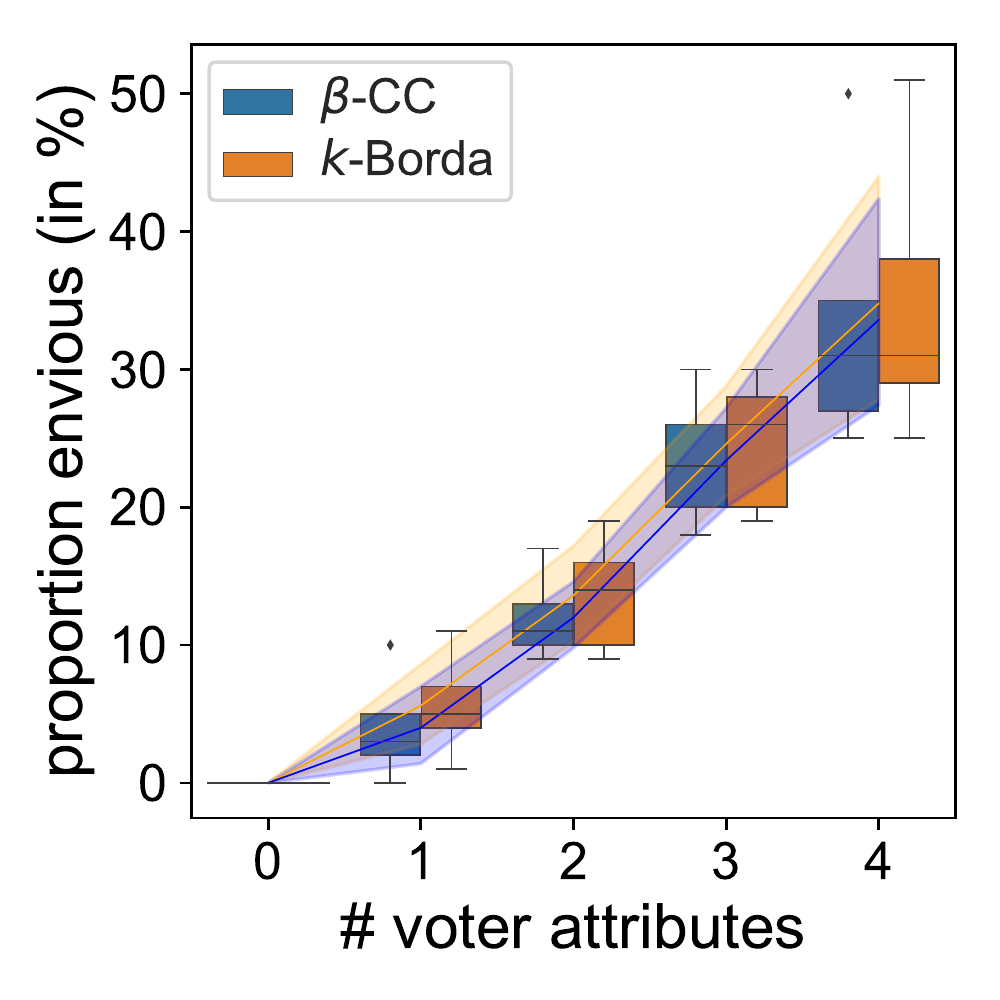}
  \caption{\textbf{FEC violated}}
  \label{fig:fec/proportion}
\end{subfigure}%
\begin{subfigure}{.23\textwidth}
  \centering
  \includegraphics[width=\linewidth, trim=0cm 0cm 0cm 0cm]{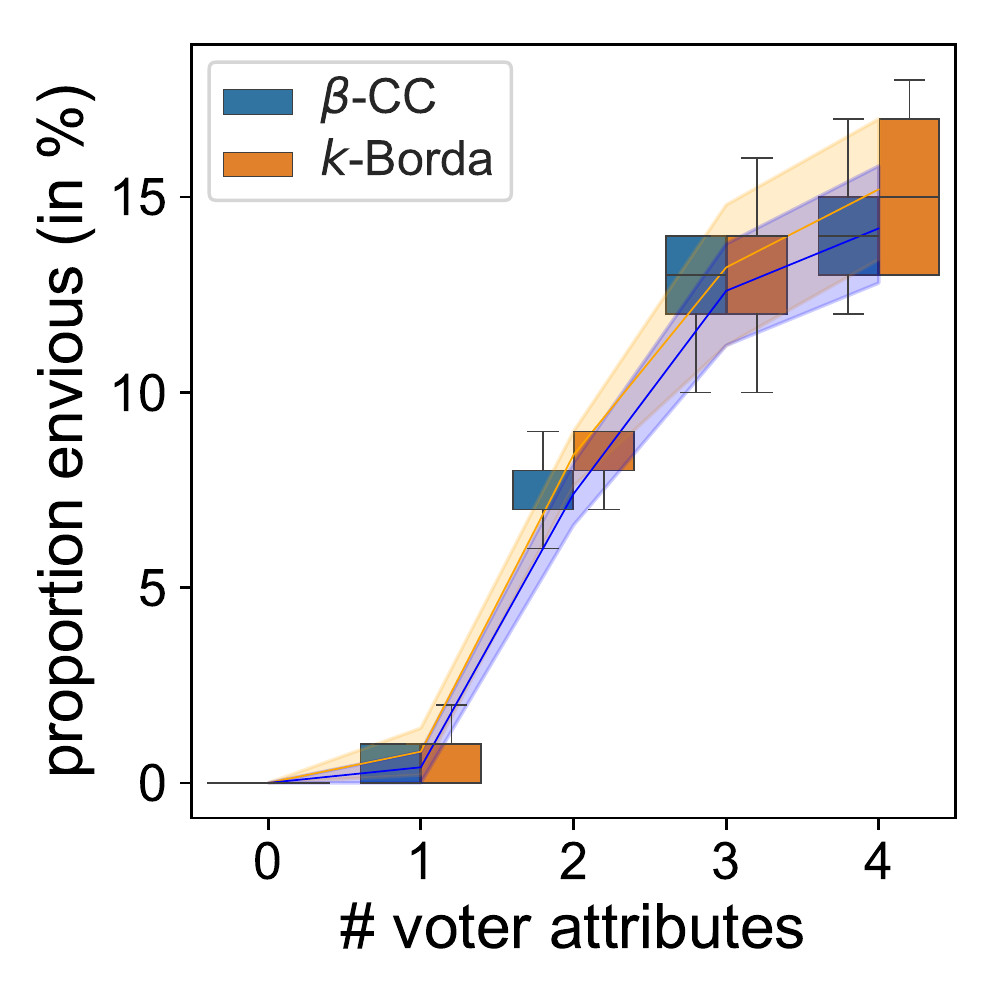}
  \caption{\textbf{WEC violated}}
  \label{fig:wec/proportion}
\end{subfigure}%
\begin{subfigure}{.23\textwidth}
  \centering
  \includegraphics[width=\linewidth, trim=0 0cm 0cm 0cm]{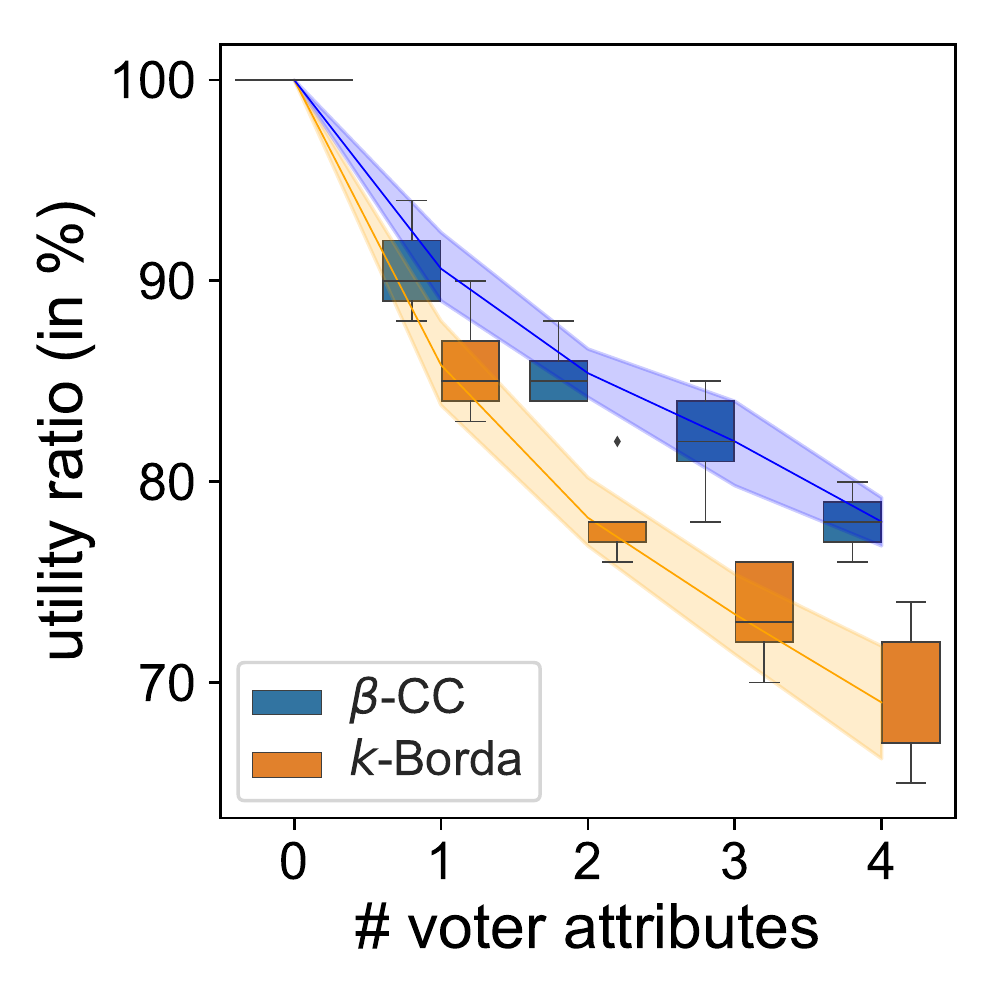}
  \caption{\textbf{Loss in utility- FEC}}
  \label{fig:fec/utility}
\end{subfigure}
\begin{subfigure}{.23\textwidth}
  \centering
  \includegraphics[width=\linewidth, trim=0 0cm 0cm 0cm]{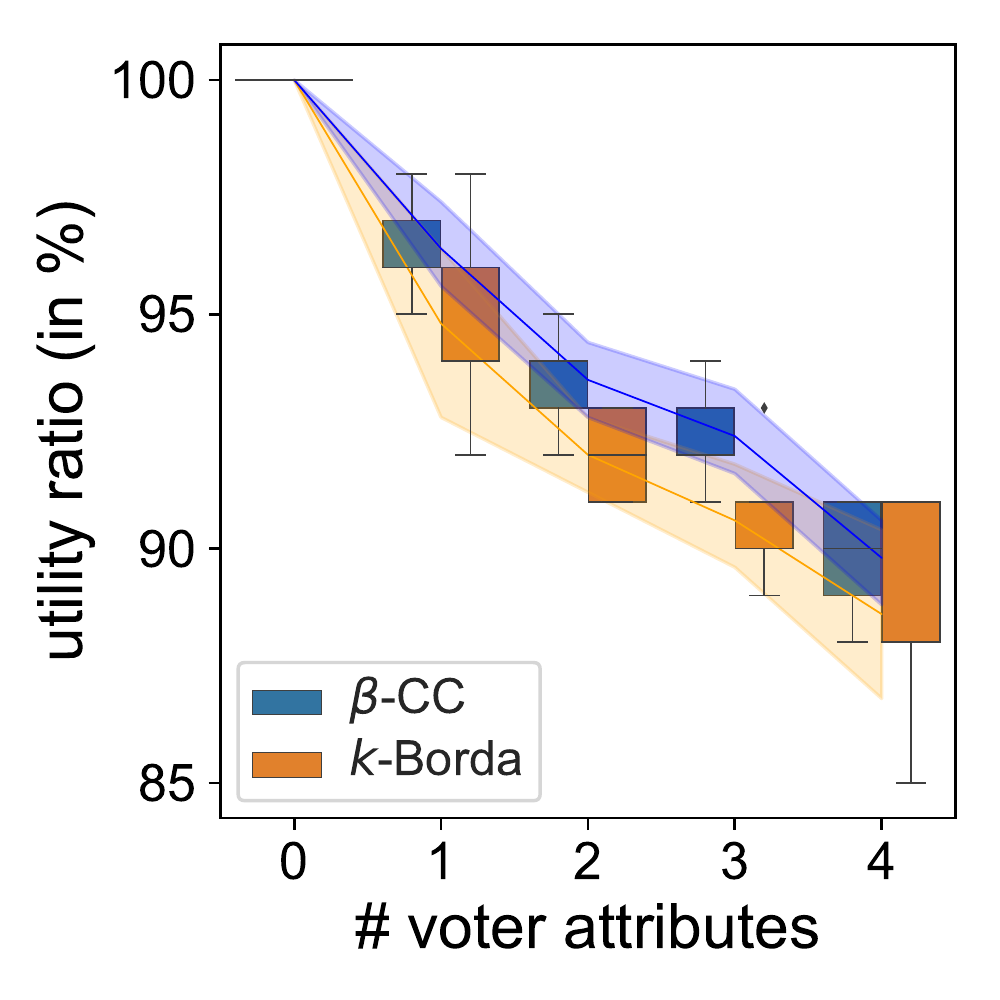}
  \caption{\textbf{Loss in utility - WEC}}
  \label{fig:wec/utility}
\end{subfigure}

\caption{The proportion of the population who are envious of at least one other population based on the criteria of (a) favorite-envyfreeness (FEC) and (b) weighted-envyfreeness (WEC). The loss in utility, in terms of the ratio of the utility of (c) FEC committee vs optimal \DiRe committee, to achieve FEC, whenever possible and (d) WEC committee vs optimal \DiRe committee, to achieve WEC, whenever possible. Each selection of $\pi$ (\# voter attributes) has 5 instances generated using \textbf{SynData1}.}
\label{fig:fec}
\end{figure*}

\section{Empirical Analysis}
\label{sec:what/empres}

We empirically assess the effect of having each version of the envyfree committee on the utility of the winning committee across different scoring rules.

\subsection{Datasets and Setup}
\label{sec:empres/data}


\paragraph{RealData 1:} The \emph{\underline{United Nations Resolutions}} dataset \cite{DVN/LEJUQZ_2009} consists of 193 UN member countries 
voting for 1681 resolutions presented in the UN General Assembly from 2000 to 2020. For each year, we aim to select a 12-sized \DiRe committee. Each candidate has two attributes, the topic of the resolution and whether a resolution was a significant vote or not. Each voter has one attribute, the continent.

\paragraph{SynData 1:} We set committee size ($k$) to 6 for 100 voters and 50 candidates. We generate complete preferences using RSM by setting selection probability $\Pi_{i, j}$ to replicate Mallows' \cite{Mallows1957} model ($\phi=0.5$, randomly chosen reference ranking $\sigma$ of size $m$) (Theorem 3, \cite{chakraborty2020algorithmic}) and preference probability $p(i)=1$, $\forall i \in [m]$. We randomly divide the candidates and voters into groups and populations, respectively.


\paragraph{SynData 2:} We use the same setting as SynData 1, except we fix $\mu$ and $\pi$ each to 2 and vary the cohesiveness of voters by setting selection probability $\Pi_{i, j}$ to replicate Mallows' \cite{Mallows1957} model's $\phi$ $\in$ $[0.1$, $1]$, with increments of 0.1.  We divide the candidates into groups and voters into populations in line with \textbf{SynData 1}.


\paragraph{System.}
We used a controlled virtual environment using Docker(R) on a 2.2 GHz 6-Core Intel(R) Core i7 Macbook Pro(R) @ 2.2GHz with 16 GB of RAM running macOS Big Sur (v11.1). 
We used Python 3.7. Note that we used a personal machine without requiring any commercial tools.

\paragraph{Constraints.}
For each $G \in \mathcal{G}$, 
we choose $l_G^D = 1$. For each $P \in \mathcal{P}$, 
we choose $l_P^R = \min(1,$ $k \cdot \frac{|P|}{n})$. Thus, each population is guaranteed at least one member, even if their size is small.

\paragraph{Voting Rules.} We use $k$-Borda and $\beta$-CC.

\paragraph{Metrics.} 
We use two metrics: (i) we compute the ratio of utilities of an envyfree committee to an optimal \DiRe committee. A higher ratio is desirable as it means a lesser difference between the two utilities. (ii) we measure the mean of the minimum proportion of populations that always remain envious in the case of FEC and the mean of the maximum difference in utilities that always remain between optimal \DiRe committee and UEC (or WEC).

\subsection{Results}

\subsubsection{Using SynData1}

\paragraph{FEC.} We found that the proportion of populations that are envious increases with an increase in the number of attributes (Figure~\ref{fig:fec/proportion}). Interestingly, we observed that an optimal \DiRe committee found using the $\beta$-CC rule was also FEC whenever the number of populations $|\mathcal{P}|$ was less than committee size $k$. A similar observation didn't hold for $k$-Borda. Similarly, the loss in utility (Figure~\ref{fig:fec/utility}) when using $\beta$-CC was lower than $k$-Borda. We note that loss in utility is inversely proportional to the utility ratio. Both these observations can be attributed to $\beta$-CC's design that ensures proportional representation.


\paragraph{UEC.} We found the proportion of populations that are not envyfree to be consistently $\geq 93\%$ for all instances, starting from voters having just one attribute. Thus, measuring the loss in utility is also not possible. This can be attributed to the strict requirement of each population having equal utility from the committee. 

\paragraph{WEC.} This metric provided the best results. While we report similar observations as that for FEC, we note a significant drop in the proportion of populations that are envious (Figure~\ref{fig:wec/proportion}) and a significant rise in the utility ratio (Figure~\ref{fig:wec/utility}), meaning the loss in utility decreased significantly to up to 15\% only.


\subsubsection{Using SynData2}
\paragraph{FEC.} We found that the proportion of populations that are envious decreases with an increase in the cohesiveness of voters (Figure~\ref{fig:fec/proportionphi}).  Similarly, the loss in utility (Figure~\ref{fig:fec/utilityphi}) when using $\beta$-CC was lower when voters were less cohesive and higher when more cohesive. We note that loss in utility is inversely proportional to cohesiveness.


\paragraph{UEC.} The proportion of populations that are not envyfree to be consistently $\geq 89\%$ for all instances, starting from voters having no cohesiveness. Thus, measuring the loss in utility is also not possible. In line with our observation previously, this can be attributed to the strict requirement of each population having equal utility from the committee. For $\eta$ relaxation as well, our previous observation of relaxation holds as UEC is better than the relaxation of FEC.

\paragraph{WEC.} This metric again provided the best results. We report similar observations as that for FEC. Promisingly, there was a rise in the utility ratio (Figure~\ref{fig:wec/utilityphi}) but a drop in the proportion of populations that are envious (Figure~\ref{fig:wec/proportionphi}). Overall, when compared to FEC, WEC turns out to be a better metric. Also, consistent with observations made for FEC, the proportion of envious populations decreased with an increase in cohesiveness and the drop utility ratio also decreased with an increase in cohesiveness. Finally, we found that the proportion of populations that are envious decreases with an increase in $\zeta$ (WEC up to $\zeta$). The decrease was steeper when $\phi$ was low (mean rate of decrease 1.5\% per $\zeta$ for $\phi=0.0$ to 0.1\% per $\zeta$ for $\pi=0.8$).

\subsubsection{Localized Fair Allocation}
Localized fair allocation was easier to satisfy as compared to global fair allocation. This observation, as expected, was coherent across both the datasets, all the three envyfreeness techniques, and their corresponding relaxations. The mean utility ratio for the non-relaxed techniques was significantly higher (94\% vs 85\% for global fair allocation). Specifically, the mean utility ratio was much higher for SynData2 vs SynData1 (98\% vs 92\%). Interestingly, there was no statistically significant relationship between the utility ratio and the mean number of population per attribute (Pearson's correlation, $r=-0.85$, $p>0.05$). This can be because of the varying number of attributes that the dataset has.

\subsubsection{Inter-sectional Fair Allocation and Simpson's Paradox}
Inter-sectional fair allocation was almost as difficult to satisfy as compared to global fair allocation. This observation was coherent across both the datasets, all the three envyfreeness techniques, and their corresponding relaxations. However, an interesting observation was the presence of instances that were not globally fairly allocated but were inter-sectionally fairly allocated. For example, a committee that was not fairly allocated between, say, females and African-Americans was fairly allocated to African-American females. We attribute this observation to Simpson's paradox. The proportion of instances where we observed the presence of Simpson's paradox was 14.8\%. 



\subsubsection{Using Real Data} 
For each year, we implemented each of our 3 models using 3 sets of constraints. Across all the years, the mean ratio of utilities of FEC to \DiRe committees was 0.88 (sd = 0.04), UEC to \DiRe committees was 0.48 (sd = 0.39), and WEC to \DiRe committees was 0.94 (sd = 0.04). As there is only one voter attribute, global fair allocation and localized fair allocation are equivalent. Next, due to Simpson's paradox, we observed that voting on  economic-development-related resolutions that was ``fair'' for all the  continents and all the economic groups of countries was not ``fair'' for economically-underdeveloped countries of Africa.  

\section{Conclusion and Future Work}
\label{sec:conc}

Our work motivates the need to fairly allocate utility in constrained multiwinner elections. 
Such analysis should also be carried out for all the actors of any system that guarantees fairness through the use of constraints. Application includes in domains such as machine learning \cite{golz2019paradoxes} and recommender systems \cite{zehlike2022fairness}.   
Next, on the technical front, we did an extensive complexity analysis. While the hardness results in the paper may seem negative, 
we expect committee size to be small in the real world. Hence, all our hardness results trivially become (fixed-parameter) tractable.

\section*{Acknowledgments}
I am indebted to Julia Stoyanovich for her generous guidance. I acknowledge the efforts of high-school students Eunice Son and Afifa Tanisa, and of undergraduate student Lauren Kirshenbaum, with the compilation and analysis of the United Nations dataset, respectively.


\bibliography{references_original}


\appendix
\vspace{0.5cm}

\hspace{2cm}{\Huge{Appendix}}

\vspace{0.5cm}
\section{Extended Related Work}

\paragraph{Fairness in Ranking and Set Selection.}
The existence of algorithmic bias in multiple domains is known \cite{baeza2016data, bellamy2018ai, celis2017ranking, danks2017algorithmic, hajian2016algorithmic, lambrecht2019algorithmic}.
The study of fairness in ranking and set selection, closely related to the study of multiwinner elections, use constraints in algorithms to mitigate bias caused against historically disadvantaged groups. Stoyanovich \etal~\shortcite{stoyanovich2018online} use constraints in the streaming set selection problems, and Yang and Stoyanovich~\shortcite{yang2017measuring} and Yang \etal~\shortcite{yang2019balanced} use them in ranked outputs. 
Kuhlman and Rundensteiner~\shortcite{Kuhlman2020rank} focus on fair rank aggregation and Bei~\etal \shortcite{bei2020candidate} use proportional fairness constraints.

\paragraph{Two-sided Fairness.}
The need for fairness from the perspective of different stakeholders of a system is well-studied. 
For instance, Patro \etal \shortcite{patro2020fairrec}, Chakraborty \etal \shortcite{chakraborty2017fair}, and Suhr \etal \shortcite{suhr2019two} consider two-sided fairness in two-sided platforms and Abdollahpouri \etal \shortcite{abdollahpouri2019multi} and Burke \shortcite{burke2017multisided} shared desirable fairness properties for different categories of multi-sided platforms\footnote{A two-sided platform is an intermediary economic platform having \emph{two distinct user groups} that provide each other with network benefits such that the decisions of each set of user group affects the outcomes of the other set \cite{rysman2009economics}. For example, the credit cards market consists of  cardholders and merchants and health maintenance organizations market consists of patients and doctors.}. However, this line of work focuses on multi-sided fairness in multi-sided platforms, which is technically different from an election. An election can be considered a ``one-sided platform'' consisting of more than one stakeholder as during an election, candidates do not make decisions that affect the voters and the set of candidates $C$ is (usually) a strict subset of voters $V$. Hence, $\delta$-sided fairness in a one-sided platform is also needed where $\delta$ is the number of distinct user-groups on the platform. More generally, $\delta$-sided fairness in $\eta$-sided platform warrants an analysis of $\delta \cdot \eta$ perspectives of fairness, i.e., the effect of fairness on each of the $\delta$ stakeholders for each of the $\eta$ fairness metrics being used. 

\section{Examples Summarized in Table~\ref{tab:FECUECWEC-examplesummary}}
\subsection{FEC}
\begin{example}
\label{eg-FEC}
The highest-scoring \DiRe committee selected in Example~\ref{eg-dire} was $W'=\{c_1, c_6, c_3, c_8\}$ ($\mathtt{f}(W')=286$). Note that this outcome fails to select Illinois' highest-ranked candidate ($c_5$), but selects California's highest-ranked candidate ($c_1$). Therefore, $W'$ is not FEC.
\begin{itemize}
    \item \textbf{FEC:} If we select $W''=\{c_1, c_5, c_3, c_8\}$ ($\mathtt{f}(W'')=285$), then both the states get their most preferred candidate. Thus $W''$ becomes FEC at a small cost of the total committee utility.
    \item \textbf{FEC up to $x$:} If $x=1$, then note that $W'$ itself is FEC up to 1 as one of the top-($x+1$) candidates of both the populations is on the committee.
\end{itemize}
 
\end{example}

\subsection{UEC}
\begin{example}
\label{eg-UEC}
The highest-scoring \DiRe committee selected in Example~\ref{eg-dire} was $W'=\{c_1, c_6, c_3, c_8\}$ ($\mathtt{f}(W')=286$). Note that if we calculate the total utility derived from the committee by each state, the smaller state achieves lower utility as compared to the larger state. Illinois' (IL) utility, $U_{IL}$ = 14 ($
(7\cdot0)+(6\cdot1)+(5\cdot0)+(4\cdot1)+(3\cdot1)+(2\cdot0)+(1\cdot1)+(0\cdot0)$) versus California's (CA) utility, $U_{CA}$ = 16. Therefore, $W'$ is not UEC.

\begin{itemize}
    \item \textbf{UEC:} If we select $W''=\{c_1, c_5, c_3, c_8\}$ ($\mathtt{f}(W'')=285$), then both the states have equal utility (15). Thus, $W''$ is UEC at a small cost of the total committee utility.
    \item \textbf{UEC up to $\eta$:} If $\eta=1$, then $W''$ is UEC up to 1 as well. If $\eta=2$, then $W'$ and $W''$ both are UEC up to 2 as the absolute difference between the utility derived by the two populations is less than or equal to two.
\end{itemize}

\end{example}

\subsection{WEC}

\begin{example}
\label{eg-WEC}
In Example~\ref{eg-WECcalc}, we calculated $\WU_{IL} = \frac{10}{13}$ and $\WU_{CA}=\frac{12}{13}$ for the highest-scoring \DiRe committee $W'=\{c_1, c_6, c_3, c_8\}$ ($\mathtt{f}(W')=286$). Therefore, $W'$ is not WEC.

\begin{itemize}
    \item \textbf{WEC:} If we select $W''=\{c_1, c_5, c_3, c_7\}$ ($\mathtt{f}(W'')=284$), then both the states have equal $\WU_{P}$ ($\frac{12}{13}$). Thus $W''$ becomes WEC at a small cost of the total committee utility.
    \item \textbf{WEC up to $\zeta$:} If $\zeta=\frac{1}{13}$, then $W''$ is WEC up to $\frac{1}{13}$ as well. If $\zeta=\frac{2}{13}$, then $W'$ and $W''$ both are WEC up to $\frac{2}{13}$ as the absolute difference between the weighted utility derived by the two populations is less than or equal to $\frac{2}{13}$.
\end{itemize}
\end{example}

\section{Details on Variants of Fair Allocation Model}
For the discussion in this section, we consider the following example:

\begin{example}
\label{eg:variantsFairAllocation}
Let an election $E$ consist of candidates $C$ and voters $V$ where voters are divided into four populations `African-American',`Caucasian',`female',`male' under two attributes `race' and `gender'.
\end{example}

\subsection{Global Fair Allocation}
In global fair allocation, we do a pairwise comparison between all populations $P, P' \in \mathcal{P}$ such that $P \neq P'$. The comparison is independent of the attribute a population belongs to.

In Example~\ref{eg:variantsFairAllocation}, the election is said have a global fair allocation if:
\begin{itemize}
    \item females are envyfree from males, Caucasians, and African-Americans
    \item males are envyfree from females, Caucasians and African-Americans
    \item African-Americans are envyfree from females, males, and Caucasians
    \item Caucasians are envyfree from females, males, and African-Americans
\end{itemize}

\subsection{Localized Fair Allocation}
In localized fair allocation, we do a pairwise comparison between all populations $P, P' \in \mathcal{P}$ such that $P \neq P'$ and both populations $P$ and $P'$ fall under the \emph{same} attribute.

Our notion of localized fair allocation is motivated by Abebe \etal \shortcite{abebe2016fair} discussion on local envy-freeness in the context of the classic cake-cutting problem. We discuss the notion of localized fair allocation of fairness in multiwinner elections. Specifically, we say that an allocation is localized if no population envies another population of the same attribute. 

Based on Example~\ref{eg:variantsFairAllocation}, consider that an allocation exists such female voters do not envy male voters and vice versa, African-American voters do not envy Caucasian voters and vice versa, but say, male voters envy Caucasian voters in every possible allocation. Here, we have a localized fair allocation but not a global fair allocation\footnote{Throughout the paper, fair allocation means global fair allocation. We always use the term \emph{localized} when referring to localized fair allocation.}. More specifically, in Example~\ref{eg:variantsFairAllocation}, the election is said have a localized fair allocation if:
\begin{itemize}
    \item females are envyfree from males
    \item males are envyfree from females
    \item African-Americans are envyfree from Caucasians
    \item Caucasians are envyfree from African-Americans
\end{itemize}

Note that global fair allocation implies localized fair allocation but not vice versa. 

\subsection{Inter-sectional Fair Allocation}
In inter-sectional fair allocation, we do a pairwise comparison between all populations $I, I'$ where $I = P \cap P'$ for all $P, P' \in \mathcal{P}$ such that $P \neq P'$ and $P, P'$ are under different attributes $A$ and $A'$, $I' = P'' \cap P'''$ for all $P'', P''' \in \mathcal{P}$ such that $P'' \neq P'''$ and $P'', P'''$ are under different attributes $A$ and $A'$, and $I \neq I'$.

Our notion of inter-sectional fairness is derived from unfairness caused to inter-sectional populations like African-American females. For instance, in Example~\ref{eg:variantsFairAllocation}, the inter-sectional populations will be African-American females, African-American males, Caucasian females, and Caucasian males. Here, the election is said have a inter-sectional fair allocation if:
\begin{itemize}
    \item African-American females are envyfree from African-American males, Caucasian females, and Caucasian males
    \item African-American males are envyfree from African-American females, Caucasian females, and Caucasian males
    \item Caucasian females are envyfree from African-American females,  African-American males, and Caucasian males
    \item Caucasian males are envyfree from African-American females,  African-American males, and Caucasian females
\end{itemize}

Overall, the definitions of the notions remain the same as the only change is in the pairs of populations being compared. In global fair allocation, all populations are compared. In localized fair allocation, populations under the same attribute are only compared. In inter-sectional, populations that result from the intersection of each pair of populations are compared.

\section{Omitted Proofs}

\subsection{Proof for Theorem~\ref{thm:FECuptoxNP-hard}}
\begin{proof}
We give a reduction from vertex cover problem on $y$-uniform hypergraphs \cite{garey1979computers} to FEC up to $x$.

An instance of vertex cover problem on $y$-uniform hypergraphs (hint: $y=x+1$) consists of a set of vertices $Z$ = $\{z_1,z_2,\dots,z_{m}\}$ and a set of $|\mathcal{P}|$ hyperedges $S$, each connecting exactly $y$ vertices from $Z$. A vertex cover $Z' \subseteq Z$ is a subset of vertices such that each edge contains at least one vertex from $Z$ (i.e. $s\cap Z'\neq\phi$ for each edge $s\in S$). The vertex cover problem on $y$-uniform hypergraphs is to find a vertex cover $Z'$ of size at most $d$. 

We construct the FEC up to $x$ instance as follows. For each vertex $z \in Z$, we have the candidate $c \in C$. For each edge $s \in S$, we have $x+1$ most favorite candidates for each $P \in \mathcal{P}$. Note that we have $|\mathcal{P}|=|S|$. In FEC up to $x$, the requirement is that at least one of the top $x+1$ candidates from each population should be on the committee. Satisfying this requirement ensures that each population is mutually envyfree up to $x$ candidates. Thus, we set $x$ = $y-1$. This corresponds to the requirement that $s\cap Z'\neq\phi$. 

Hence, we have a vertex cover $Z'$ of size at most $d$  if and only if we have an FEC up to $x$ committee $W'$ of size at most $d$ that satisfies the requirement that at least one of the top $x+1$ candidate is in the committee, for all $P \in \mathcal{P}$. 

Note that this reduction holds for all $x \in [1,$ $k-3]$. 
\end{proof}

\subsection{Proof for Theorem~\ref{thm:UECNP-hard}}
\begin{proof}
We give a reduction from the subset sum problem \cite{garey1979computers} to UEC. An instance of subset-sum problem consists of a set of non-negative integers $X$ = $\{x_1,x_2,\dots,x_{m}\}$ and a non-negative integer $sum$. The subset sum problem is to determine if there is a subset of the given set with a sum equal to the given $sum$.

We construct the UEC instance as follows. We have $|X|$ candidates, which equals $m$ candidates. Next, we have a $m$-sized scoring vector $\mathbf{s}=\{s_1, s_2, \dots, s_m\}$ such that each $s_i \in \mathbf{s}$ corresponds to $x_i \in X$ sorted in non-increasing order based on the values. Next, we have two population $P$ and $P'$ with $l_P^R= l_{P'}^R$ and let the total utility that population $P$ gets be equal to $sum$. The utility that $P'$ gets from each candidate $c$ depends on the position of $\pos_{P'}(c)$ such that $s_1$ is the utility for  the candidate on position one, $s_2$ is the utility for the candidate on position two, and so on. In UEC, the requirement is that the utility that $P$ gets from a committee $W$ should be the same as the utility $P'$ gets from $W$. Satisfying this requirement ensures that both populations are mutually envyfree. This implies that the total utility that population $P'$ gets from $W$ should be equal to $sum$. This corresponds to the requirement that there is a subset of the given set with a sum equal to the given $sum$. 

Hence, we have a subset-sum if and only if we have a UEC $W$ giving a utility of $sum$ to population $P'$ and $P$.
\end{proof}

\subsection{Proof for Corollary~\ref{thm:UECuptoetaNP-hard}}
\begin{proof}
We know that when $\eta=0$, UEC up to $\eta$ and UEC are equivalent. Hence, given that UEC is NP-hard, UEC up to $\eta$ is also NP-hard. Moreover, our reduction in the proof of  Theorem~\ref{thm:UECNP-hard} can be slightly  modified so that it holds $\forall \eta \in \mathbb{Z}: \eta \in \bigl[0,$ $\frac{(m-1)\cdot(m)}{2}\bigr)$. More specifically, we can reduce from subset sum where the sum is within a fixed distance from the given input $sum$ instead of subset-sum.

\end{proof}

\begin{figure*}[t!]
\centering
\begin{subfigure}{.23\textwidth}
  \centering
  \includegraphics[width=\linewidth, trim=0cm 0cm 0cm 0cm]{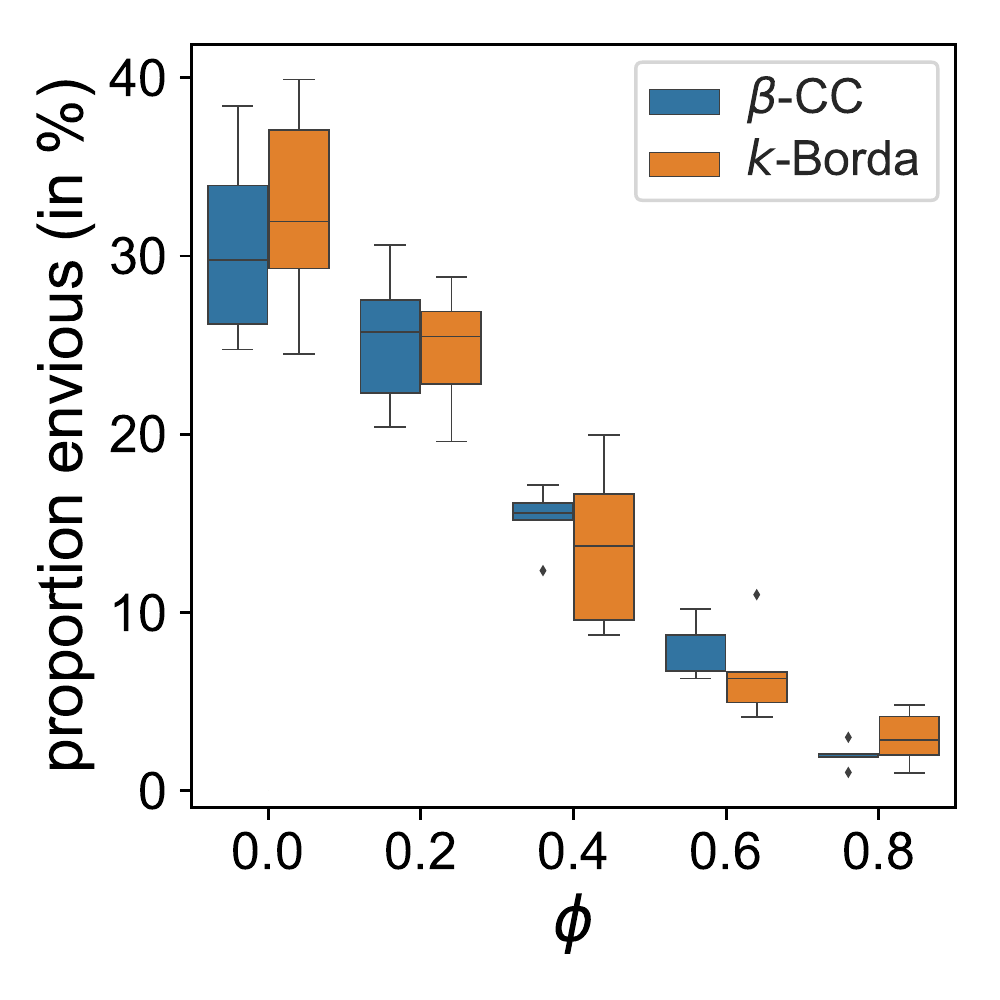}
  \caption{\textbf{FEC violated}}
  \label{fig:fec/proportionphi}
\end{subfigure}%
\begin{subfigure}{.23\textwidth}
  \centering
  \includegraphics[width=\linewidth, trim=0cm 0cm 0cm 0cm]{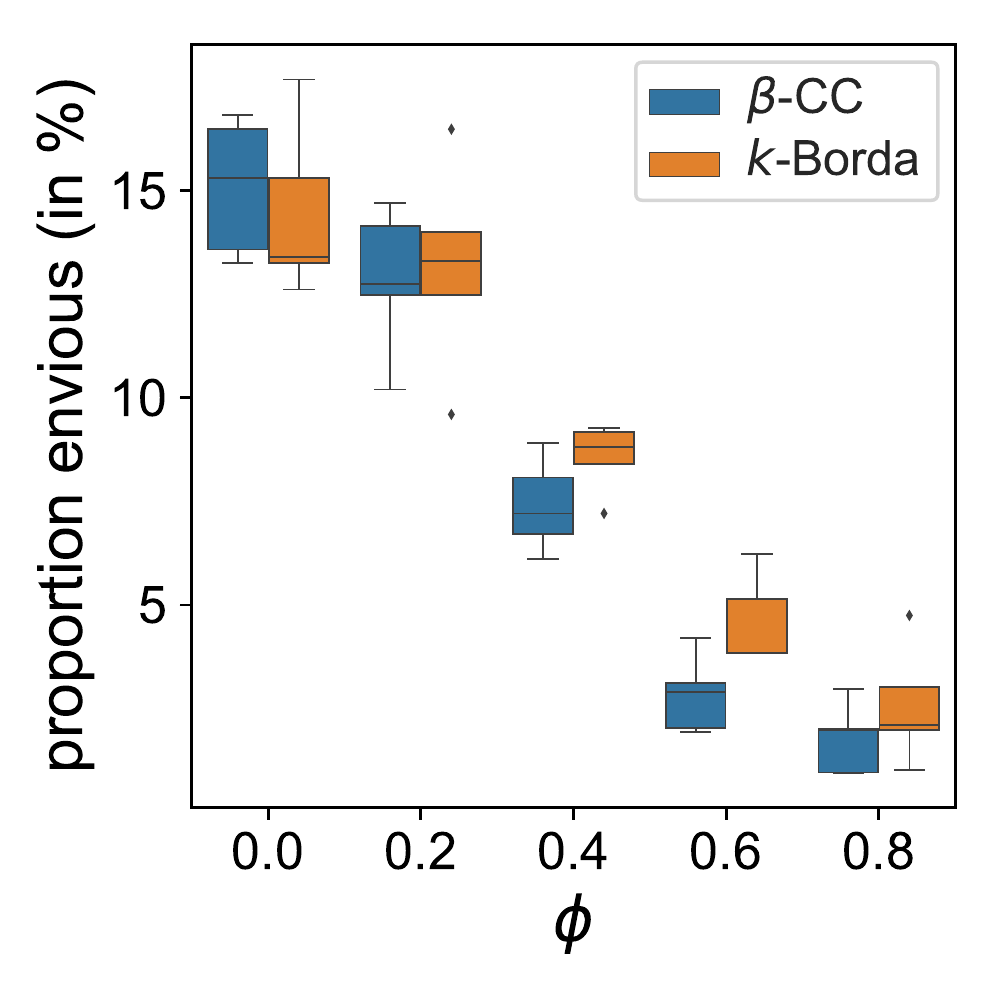}
  \caption{\textbf{WEC violated}}
  \label{fig:wec/proportionphi}
\end{subfigure}%
\begin{subfigure}{.23\textwidth}
  \centering
  \includegraphics[width=\linewidth, trim=0 0cm 0cm 0cm]{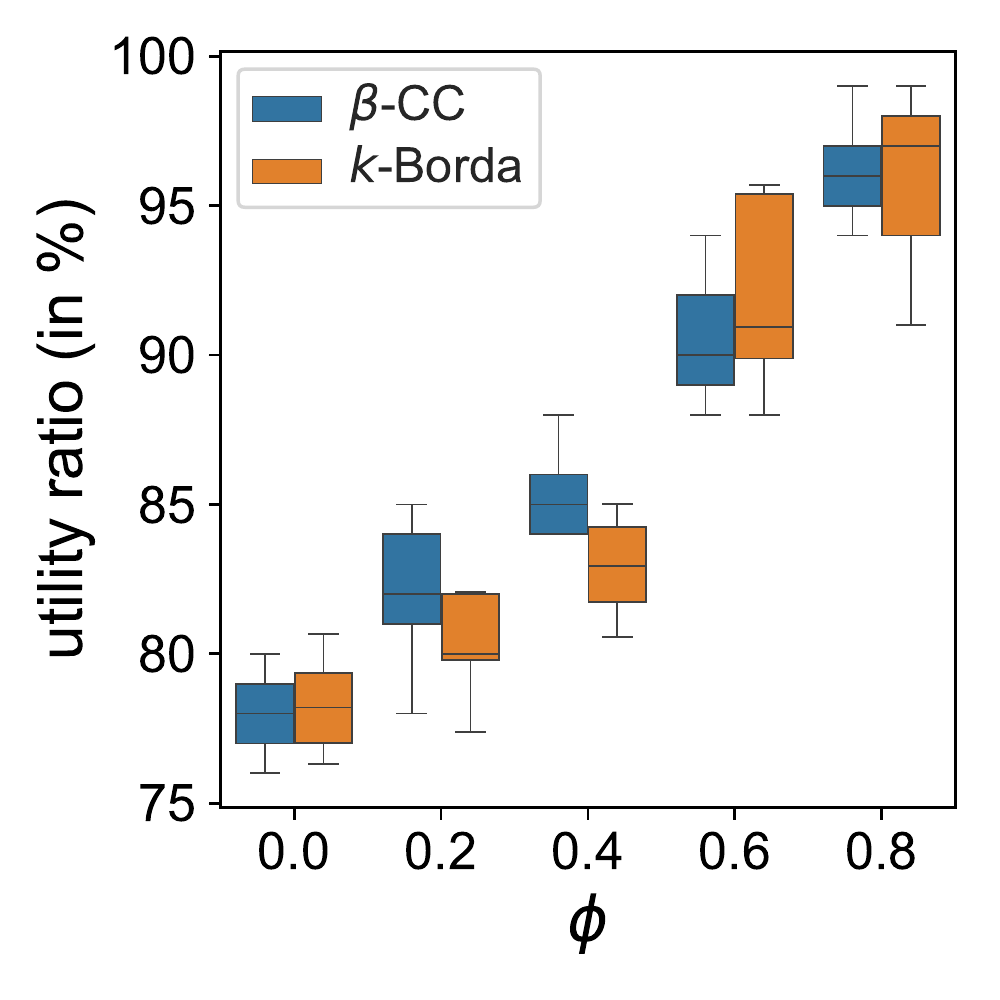}
  \caption{\textbf{Loss in utility- FEC}}
  \label{fig:fec/utilityphi}
\end{subfigure}%
\begin{subfigure}{.23\textwidth}
  \centering
  \includegraphics[width=\linewidth, trim=0 0cm 0cm 0cm]{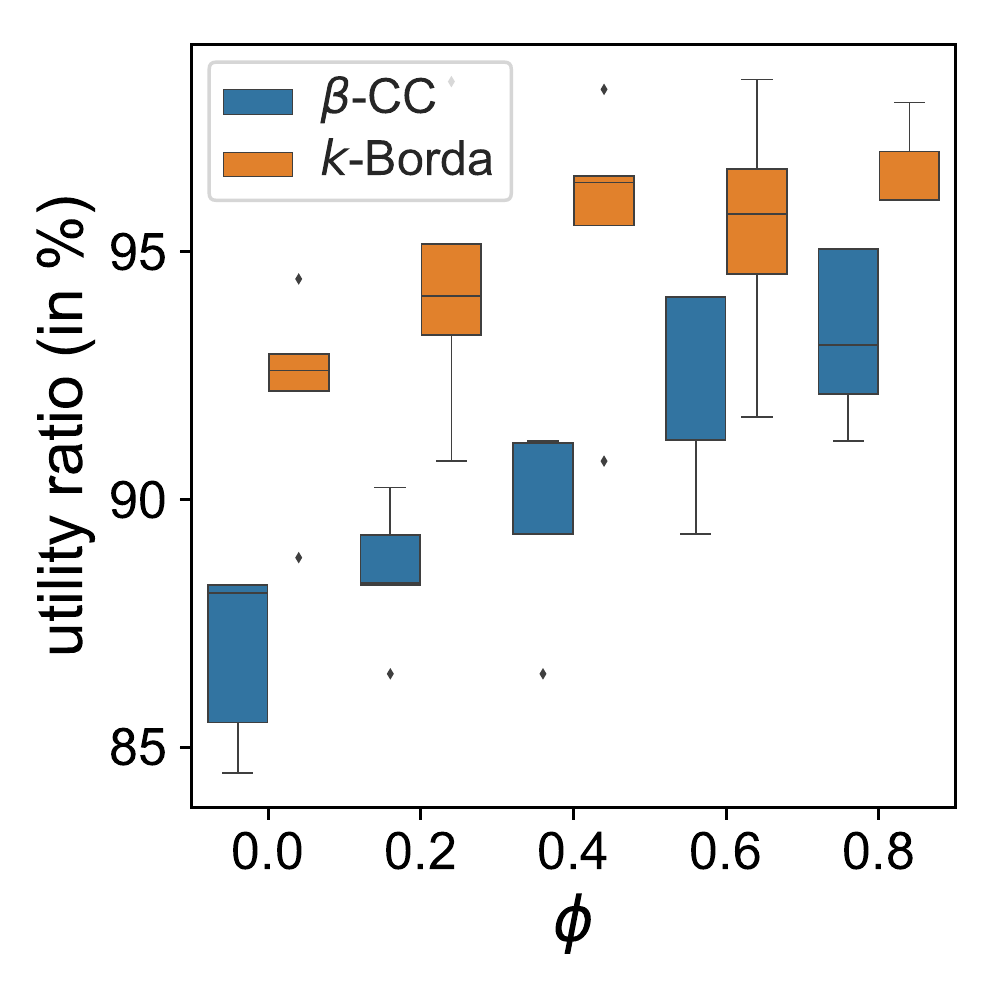}
  \caption{\textbf{Loss in utility - WEC}}
  \label{fig:wec/utilityphi}
\end{subfigure}%

\caption{The proportion of the population who are envious of at least one other population is based on the criteria of (a) favorite-envyfreeness (FEC) and (b) weighted-envyfreeness (WEC). The loss in utility, in terms of the ratio of the utility of (c) FEC committee vs optimal \DiRe committee, to achieve FEC, whenever possible and (d) WEC committee vs optimal \DiRe committee, to achieve WEC, whenever possible. Each selection of $\phi$ (voter cohesiveness) has 5 instances generated using \textbf{SynData2}.}
\label{fig:fecphi2}
\end{figure*}

\subsection{Proof for Theorem~\ref{thm:WECNP-hard}}

\begin{proof}
We give a reduction from the subset sum problem \cite{garey1979computers} to WEC. An instance of subset-sum problem consists of a set of non-negative integers $X$ = $\{x_1,x_2,\dots,x_{m}\}$ and a non-negative integer $sum$. The subset sum problem is to determine if there is a subset of the given set with a sum equal to the given $sum$.

We construct the WEC instance as follows. We have $|X|$ candidates, which equals to $m$ candidates. Next, we have a $m$-sized scoring vector $\mathbf{s}=\{s_1, s_2, \dots, s_m\}$ such that each $s_i \in \mathbf{s}$ corresponds to $x_i \in X$ sorted in non-increasing order based on the values. Next, we have two population $P$ and $P'$ such that $l_P^R= l_{P'}^R$, $W_P \cap W_{P'} \neq \phi$, and $\{\{W_P \cup W_{P'}\} \setminus \{W_P \cap W_{P'}\}\} \cap W = \phi $. Let the total utility that population $P$ gets be equal to $sum$. Hence, the weighted utility will be 
\[   \WU_P = \frac{\sum_{c \in W_P \cap W} \mathtt{f}_{\Borda}^{W_P}(c)}{\sum_{i=1}^{l_P^R}m-i} \leq \frac{\sum_{c \in W} \mathtt{f}_{\Borda}^{W_P}(c)}{\sum_{i=1}^{m}m-i} \]

Next, the $sum$ can be written as

\[sum =  \sum_{c \in W} \mathtt{f}_{\Borda}^{W_P}(c)\]

Hence, 
\[    \frac{\sum_{c \in W} \mathtt{f}_{\Borda}^{W_P}(c)}{\sum_{i=1}^{m}m-i} \leq sum  \]

Therefore, \[   \WU_P \leq sum \]

The utility that $P'$ gets from each candidate $c$ depends on the position of $\pos_P(c)$ such that $s_1$ is the utility for the candidate on position one, $s_2$ is the utility for the candidate on position two, and so on. In WEC, the requirement is that the weighted utility that $P$ gets from a committee $W$ should be the same as the weighted utility $P'$ gets from $W$. Satisfying this requirement ensures that both populations are mutually envyfree. This implies that the total utility that population $P'$ gets from $W$ should be equal to 

\[   \WU_{P'} = \frac{\sum_{c \in W_{P'} \cap W} \mathtt{f}_{\Borda}^{W_{P'}}(c)}{\sum_{i=1}^{l_{P'}^R}m-i} \]

However, as $l_P^R= l_{P'}^R$, we can rewrite $\WU_{P'}$ as 

\[   \WU_{P'} = \frac{\sum_{c \in W_{P'} \cap W} \mathtt{f}_{\Borda}^{W_{P'}}(c)}{\sum_{i=1}^{l_{P}^R}m-i} \]

Additionally, as $W_P \cap W_{P'} \neq \phi$, and $\{\{W_P \cup W_{P'}\} \setminus \{W_P \cap W_{P'}\}\} \cap W = \phi $, we know that the candidates in the winning \DiRe committee is common to both the population. Hence, we know the utilities will be equal:

\[\sum_{c \in W_{P'} \cap W} \mathtt{f}_{\Borda}^{W_{P'}}(c) = \sum_{c \in W_{P} \cap W} \mathtt{f}_{\Borda}^{W_{P}}(c)\]

Therefore, we can rewrite $\WU_{P'}$ as

\[   \WU_{P'} = \frac{\sum_{c \in W_{P} \cap W} \mathtt{f}_{\Borda}^{W_{P}}(c)}{\sum_{i=1}^{l_{P}^R}m-i} \]

This implies that 
\[ \WU_{P'} = \WU_{P}\]

However, for this equation to hold, it is necessary that $P'$  gets a total utility of $sum$ from $W$ in line with $P$  getting a total utility of $sum$. This corresponds to the requirement that there is a subset of the given set with a sum equal to the given $sum$. 

Hence, we have a subset-sum if and only if we have a WEC $W$ giving a total utility of $sum$ and weighted utility of $W_P$ to population $P'$ and $P$.
\end{proof}

\subsection{Proof for Corollary~\ref{thm:WECuptozetaNP-hard}}
\begin{proof}
We know that when $\zeta=0$, WEC up to $\zeta$ and WEC are equivalent. Hence, given that WEC is NP-hard, WEC up to $\zeta$ is also NP-hard. Moreover, our reduction in the proof of  Theorem~\ref{thm:WECNP-hard} can be slightly  modified so that it holds $\forall \zeta \in \mathbb{Q}: \zeta \in [0,$ $1)$. More specifically, we can reduce from subset sum where the sum is within a fixed distance from the given input $sum$ instead of subset-sum. $\zeta$ will be a function of this distance and $\forall P \in \mathcal{P}$, $|W_P\cap W|$ and $l_P^R$.

\end{proof}

\section{Empirical Analysis}

\subsection{Datasets and Setup}
\emph{Dividing Candidates into Groups and Voters into Populations for SynData 1 and SynData 2}
To systematically assess the impact of having each version of the envyfree committee on the utility of the winning committee, 
we generate datasets with a varying number of candidate and voter attributes by iteratively choosing a combination of $(\mu$, $\pi)$ such that $\mu$ and $\pi \in \{0,1,2,3,4\}$. For each candidate attribute, we choose a number of non-empty partitions $q \in [2$, $k]$, uniformly at random. Then to partition $C$, we randomly sort the candidates $C$ and select $q-1$ positions from $[2$, $m]$, uniformly at random without replacement, with each position corresponding to the start of a new partition. The partition a candidate is in is the attribute group it belongs to. For each voter attribute, we repeat the above procedure, replacing $C$ with $V$, and choosing $q-1$ positions from the set $[2$, $n]$. For each combination of $(\mu,\pi)$, we generate five datasets.
We limit the number of candidate groups and number of voter populations per attribute to $k$ to simulate a real-world division of candidates and voters.

\subsection{Results}
\subsubsection{SynData1}
\paragraph{FEC.}
We observed that the proportion of populations that are envious decreases with an increase in $x$ (FEC up to $x$). The decrease was gradual when $\pi$ was low (mean rate of decrease 2\% per $x$ for $\pi=2$ to 5\% per $x$ for $\pi=4$). When $x=k-1$, \DiRe committee and FEC up to $x$ are equivalent. Similar is the case with relaxations made for UEC and WEC. Hence, as all ratios converge to 100\%, we do not analyze the trend in the change in utility ratio.

\paragraph{UEC.}
Promisingly, we observed that the \emph{relaxation} of UEC up to $\eta$ was more effective than the relaxation of FEC up to $x$. Note that this does not translate to UEC being more effective than FEC. While the proportion of populations that are envious decreased with an increase in $\eta$, the decrease was steeper (mean rate of decrease of 4\% per $\eta$ for $\pi=2$ to 21\% per $\eta$ for $\pi=4$).  This is because multiple populations existed that were envious by a few points as compared to fewer populations that existed that were envious by a few candidates.

\paragraph{WEC.}
We observed that the proportion of populations that are envious decrease with an increase in $\zeta$ (WEC up to $\zeta$) (mean rate of decrease 2\% per $\zeta$ for $\pi=2$ to 0.9\% per $\zeta$ for $\pi=4$). 

\subsubsection{SynData 2}

\paragraph{FEC.}
We observed that the proportion of populations that are envious decreases with an increase in $x$ (FEC up to $x$). The decrease was steeper when $\phi$ was low (mean rate of decrease 4.0\% per $x$ for $\phi=0.0$ to 0.8\% per $x$ for $\pi=0.8$). Again, as all ratios converge to 100\%, we do not analyze the trend in the change in utility ratio.


\section{Detailed Conclusion}
The number of studies on fairness in various computer science domains is rising. Hence, it is important to ensure that enforcing fairness does not do more harm than good. This is because there is an understanding in social sciences that organizations that answer the call for fairness to avoid legal troubles or to avoid being labeled as ``racists'' may actually create animosity towards racial minorities due to their imposing nature \cite{dobbin2016diversity,bonilla2006racism, ray2019theory}. Similarly, when an election systematically is fair unequally, voters from the historically under-represented populations may feel that fairness 
comes at the cost of their representation. 
It can do more harm than good. Hence, it is important to ensure that we fairly allocate fairness. In this paper, we operationalized a model that does so by systematically assessing \emph{who} pays \emph{what} cost of fairness. We studied the computational complexity of finding a committee using each setting of our model. We also showed that manipulating \DiReCWD becomes NP-hard. We finally ran experiments to assess the effect of having each version of the envyfree committee on the utility of the winning committee across different scoring rules. We saw a direct relationship between the number of voter attributes and loss in overall utility required to balance the loss in utility across each population to have a fairer outcome.

On the technical front, a line of future work entails (i) a classification of the complexity of the model with respect to $\mu$, $\pi$, and $\mathtt{f}$, and (ii) approximation and parameterized complexity analysis of NP-hard instances.


\end{document}